\documentclass[aps,prl,superscriptaddress,twocolumn,amsfonts,amsmath,a4paper,showpacs]{revtex4}
\usepackage{graphicx}

\begin{document}

\title{Dynamic Arrest in Polymer Melts: \\
Competition between Packing and Intramolecular Barriers}

\author{Marco Bernabei}
\email[Corresponding author: ]{sckbernm@ehu.es}
\affiliation{Donostia International Physics Center, Paseo Manuel de Lardizabal 4,
20018 San Sebasti\'{a}n, Spain.}
\author{Angel J. Moreno}
\affiliation{Centro de F\'{\i}sica de Materiales, Centro Mixto CSIC-UPV/EHU, 
Apartado 1072, 20080 San Sebasti\'{a}n, Spain.}
\author{Juan Colmenero}
\affiliation{Donostia International Physics Center, Paseo Manuel de Lardizabal 4,
20018 San Sebasti\'{a}n, Spain.}
\affiliation{Centro de F\'{\i}sica de Materiales, Centro Mixto CSIC-UPV/EHU, 
Apartado 1072, 20080 San Sebasti\'{a}n, Spain.}
\affiliation{\mbox{Departamento de F\'{\i}sica de Materiales, Universidad del Pa\'{\i}s Vasco (UPV/EHU),
Apartado 1072, 20080 San Sebasti\'{a}n, Spain.}}

\begin{abstract}

We present molecular dynamics simulations of a simple model for polymer melts with intramolecular barriers. 
We investigate structural relaxation as a function of the barrier strength.
Dynamic correlators can be consistently analyzed within
the framework of the Mode Coupling Theory (MCT) of the glass transition.
Control parameters are tuned in order to induce a competition between
general packing effects and polymer-specific intramolecular barriers
as mechanisms for dynamic arrest. This competition yields unusually large values of the
so-called MCT exponent parameter and rationalize qualitatively different observations
for simple bead-spring and realistic polymers. The systematic study of the effect of intramolecular barriers
presented here also establishes a fundamental difference between the nature of the glass transition
in polymers and in simple glass-formers.

\end{abstract}
\date{\today}
\maketitle

Since they do not easily crystallize, polymers are probably the most extensively 
studied systems in relation with the glass transition phenomenon.
Having said this, their macromolecular character, and in particular chain connectivity,
must not be forgotten. Its most evident effect is the sublinear increase 
of the mean squared displacement (Rouse-like) arising after the decaging process, in contrast to 
the linear regime found in non-polymeric glass-formers.
Another particular ingredient of polymers is that, apart from fast librations or methyl group rotations, every motion, 
as local as it be, involves jumps over carbon-carbon rotational barriers and/or chain conformational changes. 

In this work we investigate, by means of molecular dynamics simulations, 
the decisive role of intramolecular barriers on the glass transition
of polymer melts, by systematically tuning barrier strength in a simple bead-spring model.
We discuss the obtained results within the framework of the Mode Coupling Theory (MCT) 
of the glass transition \cite{mctrev1}. Initially derived for monoatomic hard-sphere systems, 
the theory has been further developed for more complex systems, including fully-flexible bead-spring chains as 
simple models for polymer melts \cite{chongpre}.
MCT asymptotic laws have been tested in different polymeric systems.
The values of the associated dynamic exponents exhibit significant differences between
the limits of fully-flexible bead-spring chains \cite{baschrev} and fully-atomistic polymers \cite{narros}.
In particular, the so-called exponent parameter takes standard values $\lambda \sim 0.7$ for the former case
and values approaching the upper limit $\lambda =1$ for chemically realistic polymers \cite{narros}. 
While the former $\lambda$-values are characteristic of systems dominated by packing effects,
as the archetype hard-sphere fluid, the limit $\lambda = 1$ arises at higher-order MCT transitions \cite{schematic}.
The latter, or more generally transitions with $\lambda \lesssim 1$, arise in systems with different competing
mechanisms for dynamic arrest. These systems include short-ranged attractive 
colloids \cite{sperl,zaccarelli} (competition between short-range attraction and hard-sphere repulsion) 
or binary mixtures with strong dynamic asymmetry \cite{mixtures,krakoviack} (bulk-like caging and confinement).

Motivated by these analogies, we argue that values $\lambda \lesssim 1$ for real polymers
also arise from the competition between two distinct mechanisms for dynamic arrest: usual packing
effects and polymer-specific intramolecular barriers. Such barriers are not present in fully-flexible bead-spring chains,
which exhibit standard $\lambda$-values \cite{baschrev}. In order to shed light on this question,
we perform a systematic investigation of the interplay between packing and intramolecular barriers.
Starting from fully-flexible bead-spring chains, 
stiffness is introduced by implementing intramolecular
bending and torsion terms. The barrier strength is systematically tuned in order to induce competition
between the former two mechanisms.  We restrict to stiffness for which
no orientational order is present, and provide a complete dynamic picture of the isotropic phase
as a function of the barrier strength.
An extensive test of MCT asymptotic laws is performed. Simulation results are described with consistent
sets of MCT exponents. A progressive increase of $\lambda$ is induced by strengthening 
the competition between packing and intramolecular barriers, confirming the proposed scenario.

We simulate bead-spring chains of $N=10$ identical monomers of mass $m=1$.
Monomer-monomer interactions are given by a corrected soft-sphere potential:
$V(r) = 4\epsilon[(\sigma/r)^{12} - C_0 + C_2(r/\sigma)^{2}]$,
%
where $\epsilon=1$ and $\sigma=1$.
$V(r)$ is set to zero for  $r \geq c\sigma$, with $c = 1.15$.
The values $C_0 = 7c^{-12}$ and $C_2 = 6c^{-14}$ guarantee continuity of potential and forces at $r = c\sigma$.
$V(r)$ is purely repulsive and has no local minima. 
Thus, it drives dynamic arrest only through packing effects.
Chain connectivity is introduced through a FENE potential
\cite{baschrev}, 
%
$V_{\rm FENE}(r) = -\epsilon K_{\rm F}R_0^2 \ln[ 1-(R_0\sigma)^{-2}r^2 ]$,
%
between consecutive monomers, with $K_{\rm F}=15$ and $R_0 = 1.5$. 
We implement intramolecular barriers by means of the bending, $V_{\rm B}$, and torsion potential, $V_{\rm T}$,
proposed in Ref. \cite{bulacu} (see discussion there),
which are defined for each $i$-momomer ($1 \leq i  \leq N$) as:
$V_{\rm B}(\theta_i) = (\epsilon K_{\rm B}/2 )(\cos\theta_i - \cos\theta_0)^2$,
%
and 
$V_{\rm T}(\theta_{i},\theta_{i+1},\phi_{i,i+1}) = 
\epsilon K_{\rm T}\sin^3 \theta_{i} \sin^3 \theta_{i+1} \sum_{n=0}^3 a_n \cos^n \phi_{i,i+1}$.
%
%
Chain stiffness is tuned by varying $K_{\rm B}$ and $K_{\rm T}$.
$\theta_0=109.5^{\circ}$ and $\theta_i$ is the bending angle (for $2 \leq i \leq N-1$) 
between consecutive monomers $i-1$, $i$, and $i+1$. 
The dihedral angle $\phi_{i,i+1}$ is defined 
for the consecutive monomers  $i-1$, $i$, $i+1$, and $i+2$ ($2 \leq i \leq N-2$), as the angle between the two planes
defined by the sets ($i-1$, $i$, $i+1$) and ($i$, $i+1$, $i+2$).
The values of the coefficients $a_n$ are \cite{bulacu}:  $a_0=3.00$, $a_1=-5.90$, $a_2=2.06$, and $a_3=10.95$.

Temperature $T$, time $t$, wave vector $q$, and monomer density $\rho$ 
are given respectively in units of $\epsilon/k_B$ (with $k_B$ the Boltzmann constant), 
$\sigma(m/\epsilon)^{1/2}$, $\sigma^{-1}$, and $\sigma ^{-3}$.
We investigate, at fixed $\rho = 1$, the $T$-dependence of the dynamics for different values of the bending and torsion strength,
$(K_{\rm B}$,$K_{\rm T}) =$ (0,0), (15,0.5), (25,1), (25,4), and (35,4).
The case $(K_{\rm B}$,$K_{\rm T}) =(35,4)$ is also studied for $\rho = 0.93$.
The total number of chains is $N_{\rm c} =300$. Periodic boundary conditions are implemented.
Equations of motion are integrated in the velocity Verlet scheme \cite{frenkel}.
The system is prepared by placing the chains randomly in the simulation box, with a constraint avoiding monomer core overlap.
The initial monomer density is  $\rho=0.375$. Equilibration consists of  
a first run where the  box is rescaled
periodically by a factor $0.99< f < 1$ until the target density $\rho$ is reached, and 
a second isochoric run at that $\rho$. 
Thermalization at the target $T$ is achieved by periodic velocity rescaling.
Once the system is equilibrated, a microcanonical run is performed for production of configurations, from which 
observables are computed. For each state point, the latter are averaged over
typically 40 independent samples.

Orientational ordering (induced by chain stiffness) is discarded for all the analyzed cases 
by measuring the quantity $P_2 (\Theta) = (3\langle\cos ^2\Theta\rangle -1)/2$, 
where $\Theta$ is the angle between the end-to-end vectors of two chains, and average 
is performed over all pairs of distinct chains. In all cases we obtain negligible 
values $|P_2 (\Theta)| < 10^{-2}$.

\begin{figure}
\includegraphics[width=0.80\linewidth]{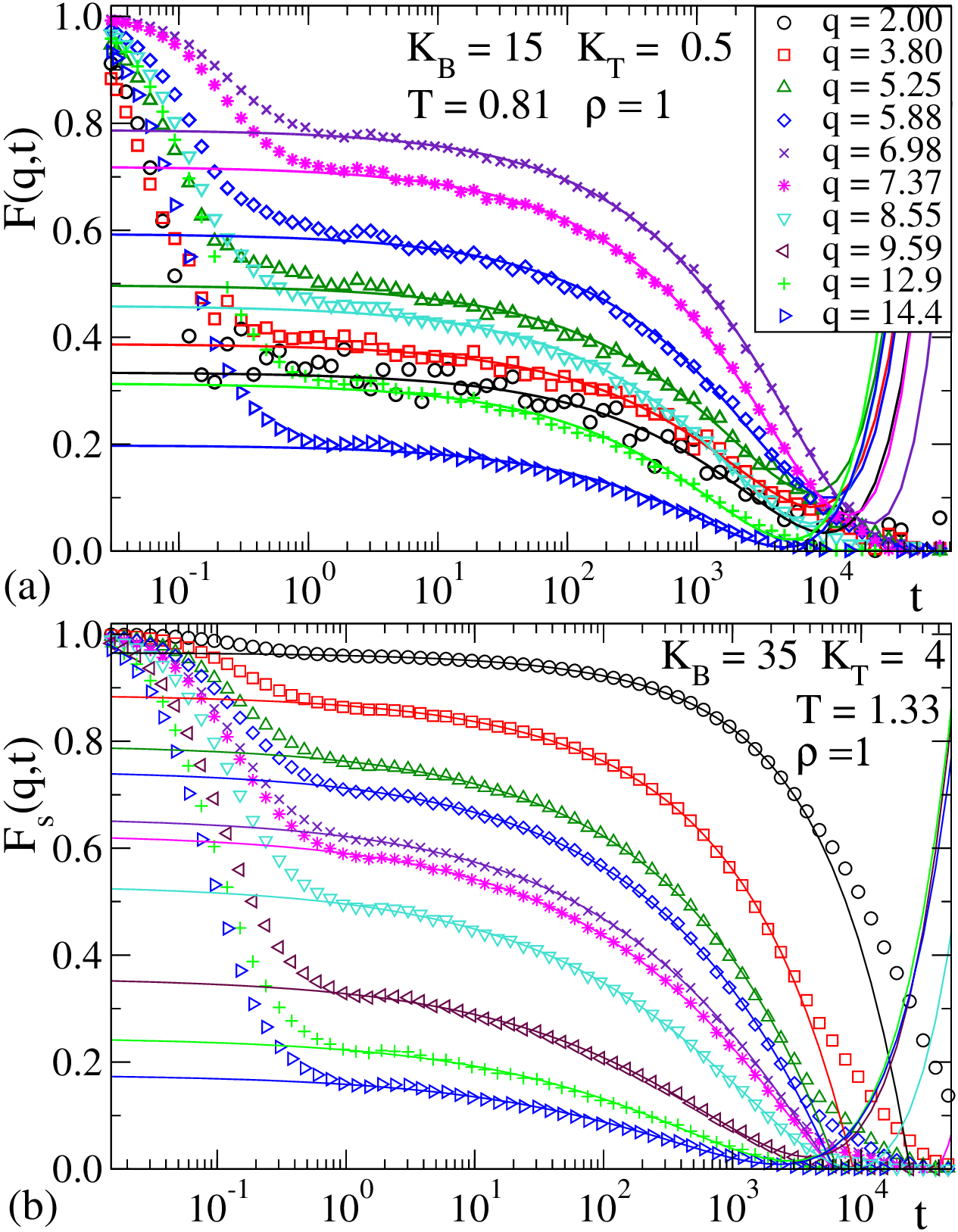}
\caption{(color online) Symbols: simulations results, at $\rho =1$, for density correlators. 
Panel (a): $F(q,t)$ for $K_{\rm B} = 15$, $K_{\rm T} = 0.5$, at $T = 0.81$. 
Panel (b): $F_{\rm s}(q,t)$ for $K_{\rm B} = 35$, $K_{\rm T} = 4$, at $T = 1.33$. 
Identical symbols in both panels correspond to identical wave vectors $q$
[values are given in panel (a)]. Lines are fits to Eq. (\ref{eqvonsch}).} 
\label{figvonsch}
\end{figure}

We compute density-density correlators, defined as 
$F(q,t)=\langle \rho(q,t)\rho(-q,0)\rangle/\langle\rho(q,0)\rho(-q,0)\rangle$
where $\rho(q,t) = \Sigma_{j}\exp[i{\bf q}\cdot {\bf r}_j(t)]$, 
the sum extending over the positions ${\bf r}_j$ of all the monomers in the system.
Density self-correlators are defined as  $F_{\rm s}(q,t) = 
(N N_{\rm c})^{-1}\Sigma_{j}\exp\{i{\bf q}\cdot [{\bf r}_j(t)-{\bf r}_j(0)]\}$.
Results for the former quantities are shown in Fig. \ref{figvonsch}, at several $q$-values,
for two state points with non-zero barriers, at $T$ close to the critical MCT temperature (see below).
As usual, a plateau is observed in the interval corresponding to the caging regime,
i.e., the temporary trapping of a particle by its neighbors.
This interval is known as the $\beta$-regime within the framework of MCT.
The second decay, corresponding to full relaxation of density fluctuations of wave vector $q$,
is known as the $\alpha$-regime, and is often described by an empirical Kohlrausch-Williams-Watts (KWW) function,
$A_q\exp[-(t/\tau^{\rm K}_q)^{\beta_q}]$, where $A_q$, the KWW time $\tau^{\rm K}_q$ 
and the exponent $\beta_q$ are $q$-dependent.

Next we summarize the basic predictions of MCT and test them in the present system. 
In its ideal version, MCT predicts a sharp transition \cite{mctrev1} from an ergodic liquid 
to an arrested state (glass) at a given value of the relevant control parameters --- here 
${\bf x}= (T,\rho,K_{\rm B},K_{\rm T})$. 
When crossing the transition point ${\bf x}={\bf x}_{\rm c}$
the long-time limit of $F(q,t)$ and $F_{\rm s}(q,t)$
jumps from zero to a non-zero value, denoted as the critical non-ergodicity parameter 
($f^{\rm c}_q$ and $f^{\rm cs}_q$, respectively). 
MCT predics asymptotic laws for dynamic observables. Such laws are characterized by dynamic exponents
that are $q$- and state-point {\it independent}. They are univoquely determined
by the static correlations at ${\bf x}={\bf x}_{\rm c}$ \cite{mctrev1}. Moreover, all the dynamic exponents
are univoquely related to a single one, the exponent parameter $\lambda$ (see below), which is
the basic one controlling all MCT asymptotic laws. Now we summarize the main ones.

For ergodic states close to ${\bf x}_{\rm c}$, the initial part of the $\alpha$-process
(i.e., the von Schweidler regime) is given by a power law expansion \cite{mctrev1}:
\begin{equation}
F(q,t) \approx f^{\rm c}_q -h_q (t/\tau_{\alpha})^{b} + h_q^{(2)}(t/\tau_{\alpha})^{2b},
\label{eqvonsch}
\end{equation}
(and analogously for self-correlators) with $0 < b \le 1$. The non-ergodicity parameters and the
prefactors $h_q$ and $h_q^{(2)}$ only depend on
$q$ and are different for each correlator. 
The $\alpha$-relaxation time $\tau_{\alpha}$ only depends on the separation parameter 
$|{\bf x} - {\bf x}_{\rm c}|$. 
MCT predicts a divergence \cite{hopping} according to the power law 
$\tau_{\alpha} \propto |{\bf x} - {\bf x}_{\rm c}|^{-\gamma}$.
In practice $\tau_{\alpha}$ can be defined as the time $\tau_{z}$ where $F(q_{\rm max},t)$ decays to some small value $z$
far below the plateau, with $q_{\rm max}$ the $q$-value at the maximum of the 
static structure factor $S(q) = (N N_{\rm c})^{-1}\langle\rho(q,0)\rho(-q,0)\rangle$. 
Here we will use $\tau_{0.2}$. The exponent $\gamma$  is given by \cite{mctrev1}: 
\begin{equation}
\gamma = (1/2a) + (1/2b).
\label{eqgamma}
\end{equation}
As mentioned above, the full $\alpha$-decay can be described by a  KWW function.
In the limit of large $q$ MCT predicts \cite{fuchs} for the KWW times a power law
$\tau^{\rm K}_q \propto q^{-1/b}$.
The exponents $a$, $b$, and $\gamma$
are univoquely related to the exponent parameter $\lambda \le 1$ through \cite{mctrev1}:
\begin{equation}
\lambda = \frac{\Gamma^{2}(1+b)}{\Gamma(1+2b)} = \frac{\Gamma^{2}(1-a)}{\Gamma(1-2a)},
\label{eqlambda}
\end{equation}
with $\Gamma$ the Euler's Gamma function.

\begin{figure}
\includegraphics[width=0.7\linewidth]{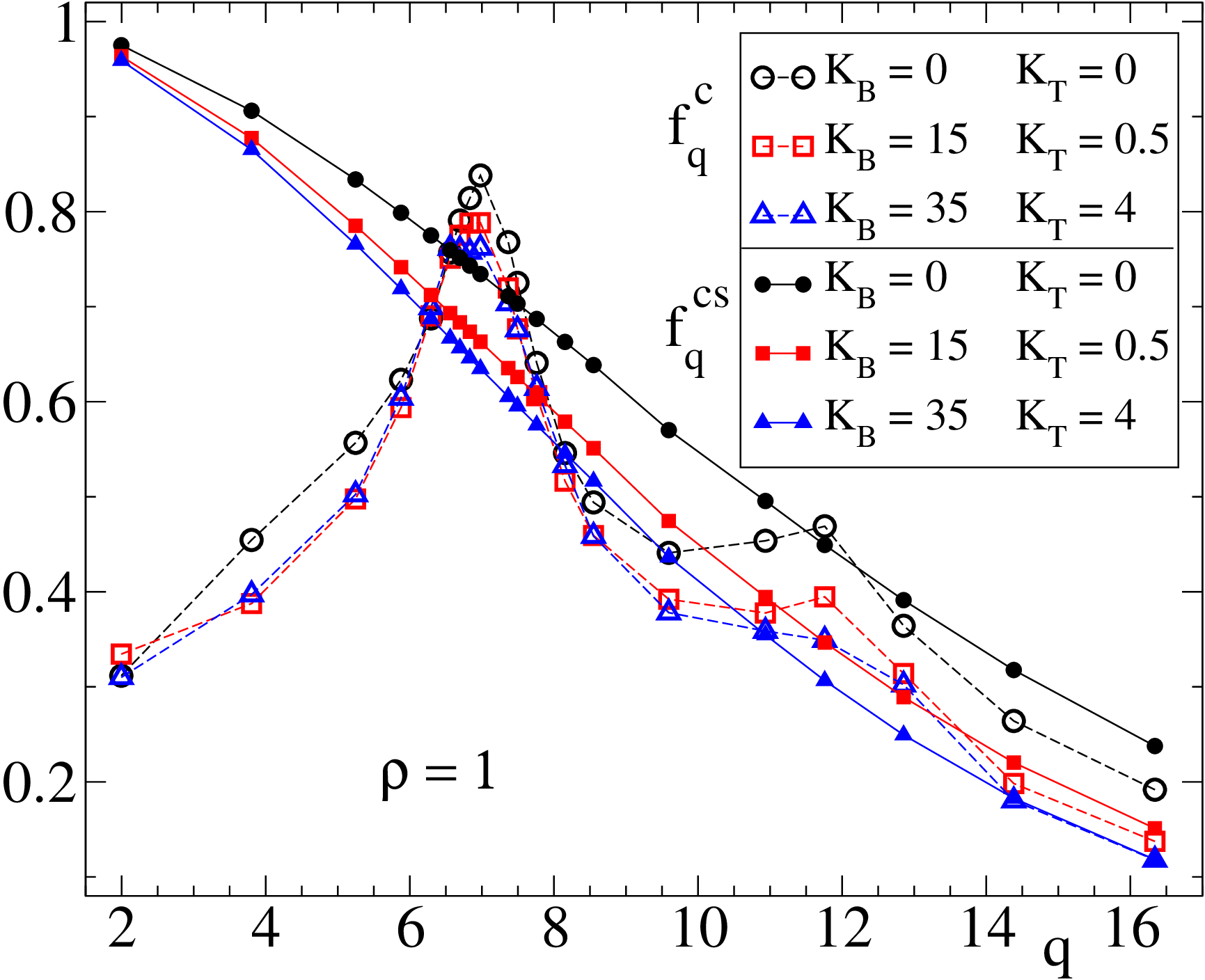}
\caption{(color online) Critical non-ergodicity parameters, as determined
from fits to Eq. (\ref{eqvonsch}), for different barrier strength at $\rho =1$.
Lines are guides for the eyes.} 
\vspace{-1 mm}
\label{figfne}
\end{figure}

\begin{figure}
\includegraphics[width=0.47\linewidth]{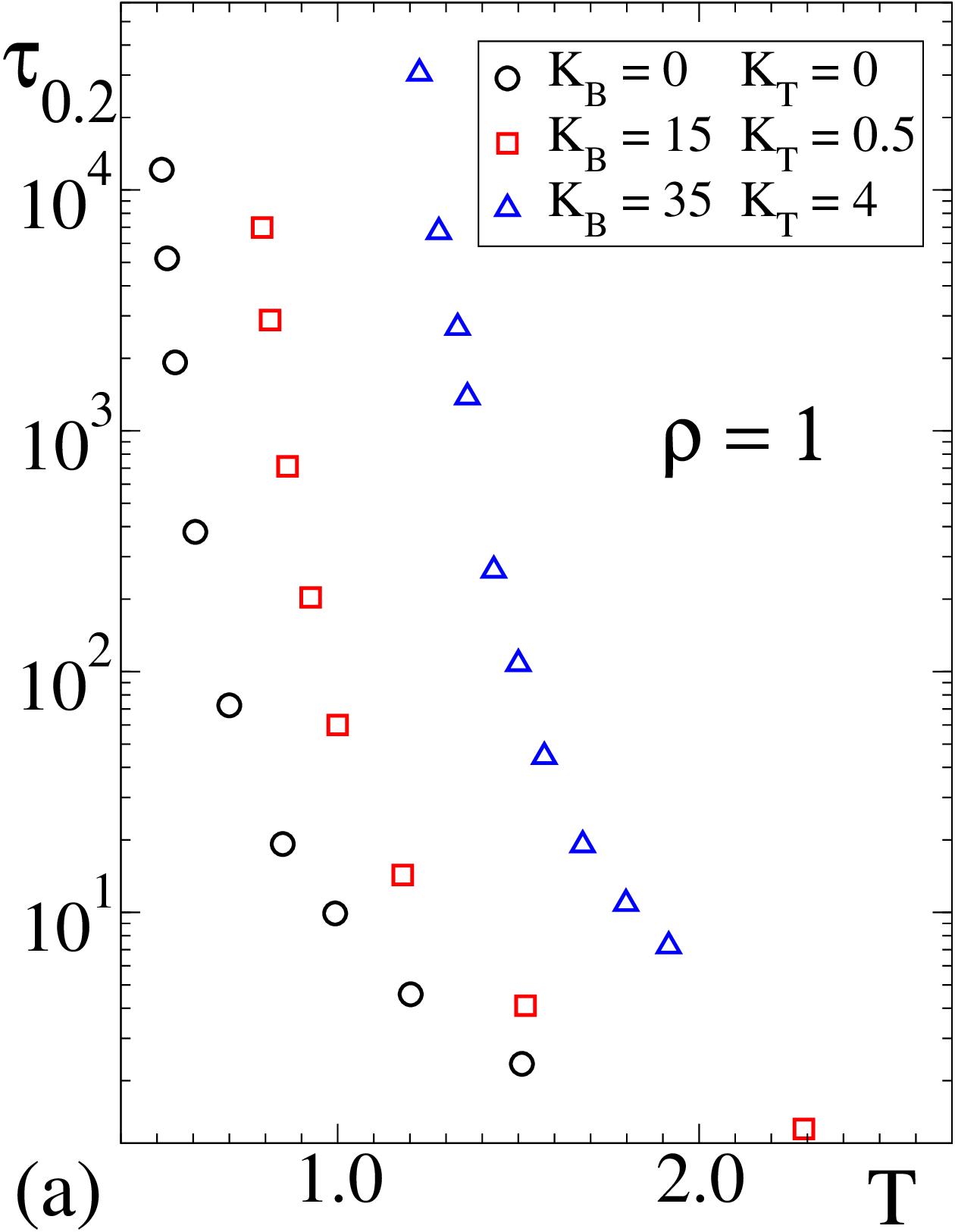}
\hspace{2 mm}
\includegraphics[width=0.47\linewidth]{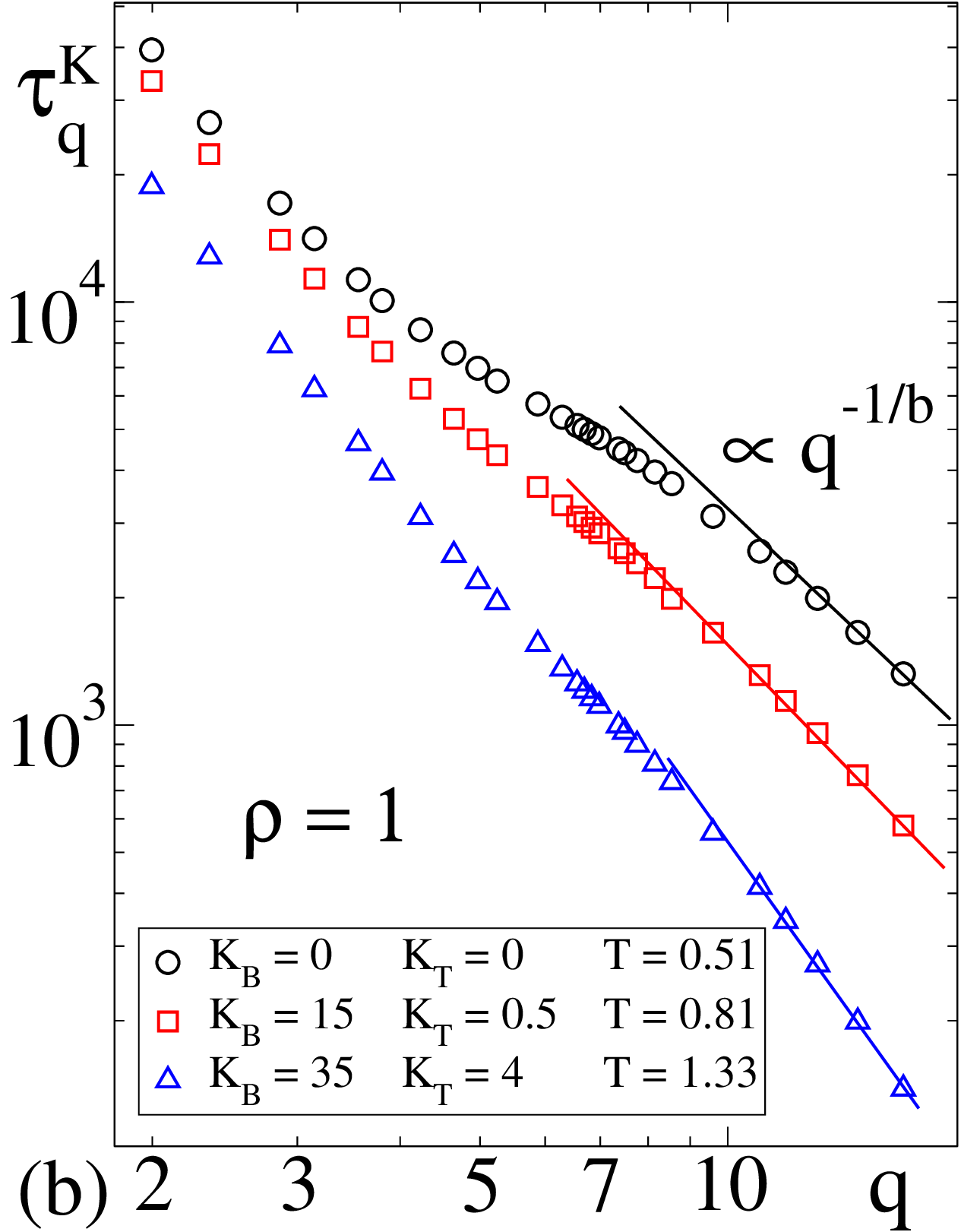}
\newline
\newline
\includegraphics[width=0.66\linewidth]{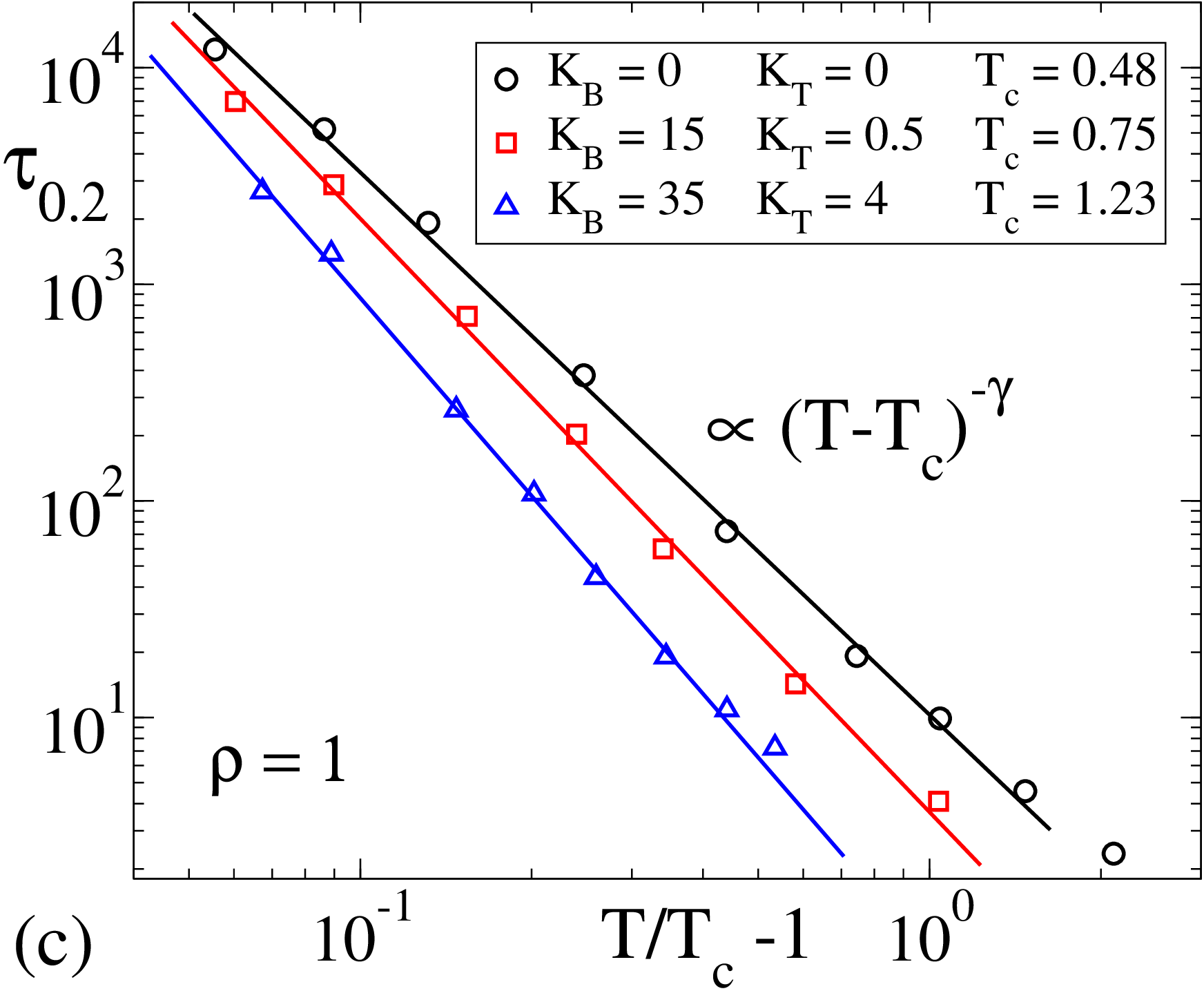}
\caption{(color online) Symbols: for $\rho=1$, $T$- and $q$-dependence of relaxation times (see text for notations) 
for different barrier strength. Lines in (b) and (c) are fits
to respectively  $\propto q^{-1/b}$ and  $\propto (T-T_{\rm c})^{-\gamma}$ (see text).} 
\vspace{-2 mm}
\label{figtau}
\end{figure}

When numerical solutions of the MCT equations are not available,
the former non-ergodicity parameters, prefactors and exponents 
are obtained as fit parameters from simulation or experimental data. 
Consistency of the analysis requires that dynamic correlators and relaxation times 
are described by a common set of exponents,
univoquely related through Eqs. (\ref{eqgamma},\ref{eqlambda}).
This consistent test has been done for all the systems here investigated, 
with different strength of the intramolecular barriers.
Figs. \ref{figvonsch}-\ref{figtau} display some representative examples.
Fig. \ref{figvonsch} shows at fixed $\rho =1$ and for a broad $q$-range, 
fits to Eq. (\ref{eqvonsch}) of density correlators
for the state points $K_{\rm B} = 15$, $K_{\rm T} = 0.5$, $T = 0.81$ (S1) 
and $K_{\rm B} = 35$, $K_{\rm T} = 4$, $T = 1.33$ (S2). A good description
is achieved, for all the $q$-values and over several time decades, with a fixed $b$-exponent 
($b = 0.50$ and 0.37 for respectively S1 and S2).
Fig. \ref{figfne} displays, for the former barrier strength, the $q$-dependence of the critical non-ergodicity 
parameters. The fully flexible case $K_{\rm B} = K_{\rm T} =0$ is also included.
As deduced from the stronger decay of 
$f_q^{\rm c}$ and $f_q^{\rm cs}$ for stronger barriers, chain stiffness induces a weaker localization
at fixed density. By making an approximate fit of $f_q^{\rm cs}$ to Gaussian behavior, 
$\exp(-q^2 l_{\rm c}^2 /6)$, we estimate, 
at fixed $\rho =1$, a localization length $l_{\rm c} =$ 0.19, 0.21, and 0.23 for respectively  
$(K_{\rm B},K_{\rm T}) =$ (0,0), (15,0.5), and (34,4).

The increase of the barrier strength at fixed $\rho$ also induces a higher critical temperature $T_{\rm c}$,
and a longer relaxation time for fixed $\rho$ and $T$. This is demonstrated in
Fig. \ref{figtau}, which also shows a test of the predictions $\tau_{\alpha} \propto (T-T_{\rm c})^{-\gamma}$
and $\tau_q^{\rm K} \propto q^{-1/b}$ (for large $q$) for the former values of the barrier strength. 
A good description is obtained with the same $b$-exponents used 
for the von Schweidler fits of Fig. \ref{figvonsch},
and with the $\gamma$-exponents derived from them through Eqs. (\ref{eqgamma},\ref{eqlambda}).
This demonstrates the consistency of the data analysis.
For comparison, Fig. \ref{figtau} also includes results for the fully flexible case
$K_{\rm B} = K_{\rm T} = 0$.

Table \ref{tab:paramkbkt} displays the results of the MCT analysis (dynamic exponents and $T_{\rm c}$)
for all the investigated cases. It also includes the mean chain end-to-end radius at $T_{\rm c}$, $R_{\rm ee}^{\rm c}$,
as computed from the simulations. $R_{\rm ee}^{\rm c}$ provides 
a qualitative characterization of chain stiffness. From numerical values in Table \ref{tab:paramkbkt} 
a clear correlation between the exponent parameter $\lambda$ and chain stiffness is unambiguously demonstrated. 
The competition between packing effects and
intramolecular barriers induces a progressive increase of $\lambda$ from the value $\lambda = 0.76$ for
fully-flexible chains to $\lambda = 0.89$ for the stiffest investigated chains.

\begin{table}
\begin{tabular}{cccccccccccccccccc}
&$\rho$ &\ &$K_{\rm B}$  &\ &$K_{\rm T}$ &\ &$R^{\rm c}_{\rm ee}$ &\ &$T_{\rm c}$ &\ &$a$  &\ &$b$  &\ &$\gamma$ &\ &$\lambda$ \\
\hline
\\
&1    &\ &0         &\ &0          &\ &3.6         &\ &0.48      &\ &0.30 &\ &0.54 &\ &2.6      &\ &0.76 \\
&1    &\ &15        &\ &0.5        &\ &5.2         &\ &0.75      &\ &0.29 &\ &0.50 &\ &2.7      &\ &0.79 \\
&1    &\ &25        &\ &1          &\ &5.5         &\ &0.82      &\ &0.26 &\ &0.43 &\ &3.1      &\ &0.83 \\
&1    &\ &25        &\ &4          &\ &6.4         &\ &1.02      &\ &0.25 &\ &0.40 &\ &3.2      &\ &0.84 \\ 
&1    &\ &35        &\ &4          &\ &6.5         &\ &1.23      &\ &0.24 &\ &0.37 &\ &3.4      &\ &0.86 \\
&0.93 &\ &35        &\ &4          &\ &6.9         &\ &1.02      &\ &0.22 &\ &0.33 &\ &3.8      &\ &0.89 \\  
\\
\hline
\end{tabular}
\caption{Values of the MCT exponents and critical temperature $T_{\rm c}$
for different $\rho$ and barrier strength. Also included
are the mean chain end-to-end radius $R_{\rm ee}^{\rm c}$ at $T_{\rm c}$.}
\label{tab:paramkbkt}
\end{table}

\begin{table}
\begin{tabular}{llcccccccc}
&System                               &\ &$a$  &\ &$b$  &\ &$\gamma$ &\ &$\lambda$ \\
\hline
\\
&Hard spheres            &\ &0.31 &\ &0.58 &\ &2.5 &\ &0.74 \\
&Orthoterphenyl              &\ &0.30 &\ &0.54 &\ &2.6 &\ &0.76 \\
&Polyethylene (UA)          &\ &0.27 &\ &0.46 &\ &2.9 &\ &0.81 \\
&1,4-Polybutadiene (UA)           &\ &0.21 &\ &0.30 &\ &4.1 &\ &0.90 \\
&1,4-Polybutadiene (FA)     &\ &0.18 &\ &0.24 &\ &4.9 &\ &0.93 \\
&Poly(vinyl ethylene) (FA)  &\ &0.18 &\ &0.24 &\ &4.9 &\ &0.93 \\
\\
\hline
\end{tabular}
\caption{MCT exponents for different glass-formers. Data are taken from \cite{narros} and references therein.
UA and FA denote respectively coarse-grained united atom  and fully-atomistic models.}
\vspace{-2 mm}
\label{tab:paramliter}
\end{table}

This observation rationalizes the large difference observed between MCT exponents for fully-flexible bead-spring
chains and chemically realistic polymers. Table \ref{tab:paramliter} shows a representative compilation
of exponents for glass-formers of very different nature. Exponents for fully-flexible bead-spring chains
are similar to those of non-polymeric glass-formers, including the hard-sphere fluid, i.e.,
the archetype glass-former dominated by packing effects. Chemically realistic polymers of increasing complexity
exhibit instead values approaching the limit $\lambda =1$ characteristic of higher-order MCT transitions.
The systematic study presented in this work strongly suggests a competition between general packing effects  
and polymer-specific intramolecular barriers as the origin of this difference. It also suggests a fundamental
difference in the nature of the glass transition in real polymers --- driven by the former competing mechanisms---
as compared to simple glass-formers \cite{notehigher}. Real polymers are thus classified in the family
of complex systems as short-ranged attractive colloids \cite{sperl,zaccarelli}
or binary mixtures with strong dynamic asymmetry \cite{mixtures,krakoviack}, 
which are characterized by an underlying higher-order MCT transition 
---or at least by unusually large values of $\lambda$---
arising from a competition between distinct mechanisms for dynamic arrest. 
Finally, results reported here provide fundamental information for microscopic theories (and in particular for MCT)
of the glass transition in polymers, which need to account for the decisive role of intramolecular barriers.

We acknowledge support from NMP3-CT-2004-502235 (SoftComp, EU), 
MAT2007-63681 (Spain), 2007-60I021 (Spain), and IT-436-07 (GV, Spain).


\begin{thebibliography}{9}


\bibitem{mctrev1} W. G\"{o}tze and L. Sj\"{o}gren, Rep. Prog. Phys. {\bf 55}, 241 (1992).
\bibitem{chongpre} S.-H. Chong {\it et al.}, 
Phys. Rev. E {\bf 76}, 051806 (2007).
\bibitem{baschrev} J. Baschnagel and F. Varnik, J. Phys.: Condens. Matter {\bf 17}, R851 (2005).


\bibitem{narros} J. Colmenero {\it et al.}, 
J. Phys.: Condens. Matter {\bf 19}, 205127 (2007);
A. Narros, Ph.D. thesis, Universidad del Pa\'{\i}s Vasco (UPV/EHU), 2007.

\bibitem{schematic} 
W. G\"{o}tze and M. Sperl, Phys. Rev. E {\bf 66}, 011405 (2002). 


\bibitem{sperl} M. Sperl, Phys. Rev. E {\bf 68}, 031405 (2003).
\bibitem{zaccarelli} E. Zaccarelli {\it et al.}, 
Phys. Rev. E {\bf 66}, 041402 (2002).


\bibitem{mixtures} A. J. Moreno and J. Colmenero, 
Phys. Rev. E {\bf 74}, 021409 (2006); J. Chem. Phys {\bf 125}, 164507 (2006).
\bibitem{krakoviack} V. Krakoviack, Phys. Rev. Lett. {\bf 94}, 065703 (2005).


\bibitem{bulacu} M. Bulacu and E. van der Giessen, J. Chem. Phys. {\bf 123}, 114901 (2005);
Phys. Rev. E {\bf 76}, 011807 (2007).
\bibitem{frenkel} D. Frenkel and B. Smit, {\it Understanding Molecular Simulation}
(Academic Press, San Diego, 1996).


\bibitem{hopping} Actually, deviations from this power law are generally observed for sufficiently small
values of $|{\bf x} - {\bf x}_{\rm c}|$. The sharp transition and time divergence are prevented
by ergodicity-restoring hopping events not included in ideal MCT \cite{mctrev1}. Still, MCT predictions
are usually fulfilled within a broad ${\bf x}$-window \cite{mctrev1}.


\bibitem{fuchs} M. Fuchs, J. Non-Cryst. Solids {\bf 172}, 241 (1994).


\bibitem{notehigher} Confirmation of an underlying higher-order transition ($\lambda =1$)
would require solving the MCT equations. Work on this question is in progress.
Having said this, the trend displayed by data of Table \ref{tab:paramkbkt} suggests that
the former would be located at regions of stronger barriers or lower $\rho$. There,
the isotropic phase becomes unstable. 
This prevents the observation of values $\lambda \rightarrow 1^{-}$.


\end{thebibliography}
\end{document}